# Bound states of gain-guided solitons in a passively mode-locked fiber laser

**L. M. Zhao, D. Y. Tang, and X. Wu**
School of Electrical and Electronic Engineering, Nanyang Technological University, Singapore
**D. J. Lei and S. C. Wen**
School of Computer and Communication, Hunan University, Changsha 410082, P. R. China

We report on the observation of bound states of gain-guided solitons (GGSs) in a dispersion-managed erbium-doped fiber laser operating in the normal net cavity dispersion regime. Despite of the fact that the GGS is a chirped soliton and there is strong pulse stretching and compression along the cavity in the laser, the bound solitons observed have a fixed pulse separation, which is invariant to the pump strength change. Numerical simulation confirmed the experimental observations.

Multipulse formation is a generic feature of the conventional soliton fiber lasers. In the case of a fiber laser mode-locked by the nonlinear polarization rotation (NPR) technique, multiple pulses are formed as a result of the cavity pulse peak clamping effect [1] and the strong pumping. The cavity feedback effect and the soliton nature of the formed pulses eventually lead to that all the pulses in cavity have identical pulse parameters. Depending on the separation between the solitons, various stable multipulse patterns can be formed. These include the random irregular pulse separations [2], harmonic mode-locking [3], soliton bunching [4-6], and multi-pulse solitons [7]. Among them the multi-pulse soliton states are of special interesting as they are formed by the direct soliton-soliton interaction [8], and exhibit fixed discrete soliton separations.

The conventional soliton fiber lasers operate in the anomalous net cavity dispersion regime, where the soliton is formed due to the balanced interaction between the anomalous cavity dispersion and the fiber nonlinear Kerr effect. In fact, solitary pulse can also be formed in fiber lasers with net normal cavity dispersion, where soliton is formed as a result of the mutual interaction among the laser gain filtering, positive cavity dispersion and the fiber nonlinear Kerr effect. As the gain filtering plays an essential role on the soliton formation in the lasers, the soliton is also called the gain-guided soliton (GGS) [9-12]. Although the GGS has different formation mechanism from the conventional solitons, it was found that multiple GGSs can still be formed in the fiber lasers, and even that the formed multiple GGSs have identical pulse parameters [12].

In this paper we further report on the experimental observation of bound states of the GGSs in a dispersion-managed fiber laser. We show that, like the conventional solitons, under direct soliton interaction the GGSs can bind together to form states of bound solitons. In addition, we also show numerically that despite of the fact that in a dispersion-managed laser the soliton circulation in cavity experiences strong pulse stretching and compression, the soliton separation in the bound solitons always remain the same value along the cavity, even due to the pulse compression the soliton separation in some segments of the cavity becomes temporally far larger than the soliton pulse width.

The fiber laser used is shown in Fig. 1. It comprises of 1.72 m erbium-doped fiber (EDF) with GVD of about -50 (ps/nm)/km and peak absorption of 150 dB/m in the 1530 nm band, standard single mode fiber (SMF) whose GVD is about 18 (ps/nm)/km, and dispersion compensation fiber (DCF) with GVD of -0.196 (ps/nm)/km. Dispersion management is achieved by changing the length of the SMF in the cavity. The NPR technique was used to achieve the self-started mode locking. To this end a polarization-dependent isolator and two polarization-controllers, one consists of two quarter-wave plates and the other two quarter-wave plates and one half-wave plate, were assembled on a fiber bench in the cavity. The laser is pumped by a 1480nm pump source and the generated pulses were output via a 10% fiber output coupler. The coupler is made of DCF and all other fibers except the gain medium are SMF.

The dispersion caused by the bulk components is negligible. With 2.8 m total length of SMF the net cavity dispersion is estimated about 0.046 $ps^2$. Self-started GGS operation was obtained in the laser through appropriately setting the orientations of the waveplates. Under strong pumping multiple GGSs were also generated in the laser. Fig. 2a-2c show for example a case of two GGSs in the cavity. The GGS feature of the pulses is evident by their characteristic steep spectral edge and the pump-dependent edge-to-edge spectral width [11]. Different from the GGSs formed in fiber lasers made of purely normal dispersion fibers [10], the short wavelength side of the optical spectrum exhibits a Gaussian-like curve, which shows the influence of the cavity dispersion-management on the GGS features. Fig. 2b shows the corresponding autocorrelation trace. It shows that the FWHM of the pulse width is about 2.06 ps if a Gaussian profile is assumed. The two GGSs are far apart in the cavity as shown in Fig. 2c, and the peak power of the GGS is measured as about 6.1 W. Experimentally we have also observed harmonic mode locking state of 3 GGSs.

The detailed soliton parameters such as the peak power, pulse width, vary with the pump power and the orientation settings of the waveplates. Through rotating the waveplates a state of bound GGSs was also experimentally obtained. Fig. 2d-2f show the bound GGSs observed. The optical spectrum of the state is shown in Fig. 2d, which has clear spectral modulations. Different from Fig. 2a, obvious spectral spikes [11] also appeared on the spectral edges. High resolution zoom-in of the spectrum is also displayed in Fig. 2d. It shows that the period of the spectral modulation is about $\Delta\lambda$=1.52 nm. Taking the central wavelength of the soliton as at 1580 nm, it corresponds to a temporal pulse spacing of about $\Delta T$=5.5ps. Fig. 2e shows the corresponding autocorrelation trace of the bound solitons. It shows three peaks with a height ratio of 1:2:1, indicating that the two GGSs are identical. Only one bound-soliton exists in the cavity as shown by Fig. 2f, and the measured soliton spacing is about 6.1 ps. The autocorrelation trace shows that the soliton pulse width is about 183 fs if a Gaussian profile is assumed, which is far narrower than the soliton pulse width measured when the GGSs are far apart in the cavity as shown in Fig. 2b. The peak power of each GGSs measured is about 89.1 W. The state of bound solitons is stable. Once it is formed, slightly changing the laser operation conditions, such as the pump strength, orientations of the waveplates, does not change the soliton separation but the soliton peak power, suggesting that they are tightly bound.

At first sight the result is out of our expectation as the ratio between the measured soliton separation and the pulse width is 34.4, which excludes the possibility of effective direct soliton-soliton interaction between the solitons [8]. To understand the formation of the bound GGSs in the fiber laser, especially their strong coupling under the large soliton separation, we conducted numerical simulations. We used the same model as reported in [11] and also the following parameters: GVD coefficients of the EDF and SMF are -50 (ps/nm)/km and 18 (ps/nm)/km, respectively, GVD of the DCF is -0.196 (ps/nm)/km, cavity length $L=0.3_{DCF}+0.3_{SMF}+1.7_{EDF}+2.5_{SMF}=4.8m$, 10% output and gain bandwidth $\Omega_g=16nm$. Stable self-started mode locking can always be obtained in the laser. Especially, with the narrow gain-bandwidth selection, multiple GGSs were generated in our simulations as the pump strength is increased.

Like the experimental observations, the soliton parameters of the numerically calculated stable GGS strongly depend on the selection of the linear cavity phase delay bias, which corresponds experimentally to change the orientations of the waveplates, and sets the level of the cavity pulse peak clamping at different levels. When the pulse peak is clamped at a low value, the formed GGS has relatively broader pulse width, while if it is at a high value, the pulse width could become significantly narrow, in addition, spectral spikes also appear on the edges of the soliton spectrum, indicating that the spikes are formed due to the strong pulse nonlinearity. Bound states of the GGSs were also numerically obtained under strong pulse nonlinearity. Fig. 3 shows for example the case obtained with the linear cavity phase delay bias setting at $1.85\pi$ and the small signal gain coefficient G=2800 [1]. We plotted the state at the cavity positions where the soliton has the maximal and the minimal pulse width. Due to the dispersion-management the pulse width of the soliton varies from about 2ps to about 200fs, while the soliton separation is fixed at the 6.3 ps.

Based on the numerical simulation it becomes clear why tightly bound solitons could still be formed in the laser even the experimentally measured soliton separation is far larger than 5 times of the soliton pulse width [8]. Obviously, as shown in Fig. 4, in other positions of the cavity, the pulse width is substantially broader than that measured at the laser output (near the position of minimum pulse width), and the soliton profiles are directly overlapped and consequently results in direct soliton interaction. The GGS is essentially a chirped soliton, along the cavity the soliton chirp also varies. Nevertheless, our numerical simulation shows that despite of the soliton chirp variation, as well as the strong pulse stretching and compression, the soliton separation between the bound solitons remains constant. Our numerical simulation further suggests that like the soliton formation in a laser is due to the mutually balanced interaction among different effects over the cavity roundtrip, the formation of the bound GGSs is a result of direct soliton interaction over the cavity.

In conclusion, bound states of GGSs have been experimentally observed in a dispersion-managed erbium-doped fiber laser operating in the normal net cavity dispersion regime. Like the case of bound soliton formation of the conventional solitons, the formation of the bound GGSs is also a result of the direct soliton-soliton interaction. In particular, our numerical study shows that it is sufficient to form tightly bound solitons if over part of the cavity round trip the soliton profiles are overlapped.

## References

1. D. Y. Tang, L. M. Zhao, B. Zhao, and A. Q. Liu, “Mechanism of multisoliton formation and soliton energy quantization in passively mode-locked fiber lasers,” Phys. Rev. A **72,** 043816 (2005).
2. D. J. Richardson, R. I. Laming, D. N. Payne, V. J. Matsas, and M. W. Philips, “Pulse repetition rates in passive, self starting, femtosecond soliton fibre laser,” Electron. Lett. **27**, 1451-1453 (1991).
3. A. B. Grudinin, D. J. Richardson, and D. N. Payne, “Passive harmonic mode-locking of a fibre soliton ring laser,” Electron. Lett. **29**, 1860–1861 (1993).
4. Boris A. Malomed, “Bound solitons in the nonlinear Schrödinger – Ginzburg-Landau equation,” Phys. Rev. A **44**, 6954-6957 (1991).
5. P. Grelu, F. Belhache, F. Gutty, and J. -M. Soto-Crespo, "Phase-locked soliton pairs in a stretched-pulse fiber laser," Opt. Lett. **27**, 966-968 (2002)
6. B. Ortaç, A. Hideur, M. Brunel, C. Chédot, J. Limpert, A. Tünnermann, and F. Ö. Ilday, “Generation of parabolic bound pulses from a Yb-fiber laser,” Opt. Express **14**, 6075-6083 (2006).
7. D. Y. Tang, W. S. Man, H. Y. Tam, and P. D. Drummond, “Observation of bound states of solitons in a passively mode-locked fiber laser”, Phys. Rev. A. **64**, 033814 (2001).
8. D. Y. Tang, B. Zhao, L. M. Zhao, and H. Y. Tam, “Soliton interaction in a fiber ring laser,” Phys. Rev. E **72**, 016616 (2005).
9. P. A. Bélanger, L. Gagnon and C. Paré, “Solitary pulses in an amplified nonlinear dispersive medium”, Opt. Lett., **14**, 943-945 (1989).

10. L. M. Zhao, D. Y. Tang, and J. Wu, "Gain-guided soliton in a positive group dispersion fiber laser," Opt. Lett. **31**, 1788-1790 (2006).
11. L. M. Zhao, D. Y. Tang, T. H. Cheng, and C. Lu, "Gain-guided solitons in dispersion-managed fiber lasers with large net cavity dispersion," Opt. Lett. **31**, 2957-2959 (2006).

L. M. Zhao, D. Y. Tang, T. H. Cheng, H. Y. Tam, and C. Lu, "Generation of multiple gain-guided solitons in a fiber laser," Opt. Lett. **32**, 1581-1583 (2007).

References

1. D. Y. Tang, L. M. Zhao, B. Zhao, and A. Q. Liu, Phys. Rev. A **72,** 043816 (2005).
2. D. J. Richardson, R. I. Laming, D. N. Payne, V. J. Matsas, and M. W. Philips, Electron. Lett. **27**, 1451 (1991).
3. A. B. Grudinin, D. J. Richardson, and D. N. Payne, Electron. Lett. **29**, 1860 (1993).
4. Boris A. Malomed, Phys. Rev. A **44**, 6954 (1991).
5. P. Grelu, F. Belhache, F. Gutty, and J. -M. Soto-Crespo, Opt. Lett. **27**, 966 (2002).
6. B. Ortaç, A. Hideur, M. Brunel, C. Chédot, J. Limpert, A. Tünnermann, and F. Ö. Ilday, Opt. Express **14**, 6075 (2006).
7. D. Y. Tang, W. S. Man, H. Y. Tam, and P. D. Drummond, Phys. Rev. A. **64**, 033814 (2001).
8. D. Y. Tang, B. Zhao, L. M. Zhao, and H. Y. Tam, Phys. Rev. E **72**, 016616 (2005).
9. P. A. Bélanger, L. Gagnon and C. Paré, Opt. Lett. **14**, 943 (1989).
10. L. M. Zhao, D. Y. Tang, and J. Wu, Opt. Lett. **31**, 1788 (2006).
11. L. M. Zhao, D. Y. Tang, T. H. Cheng, and C. Lu, Opt. Lett. **31**, 2957 (2006).
12. L. M. Zhao, D. Y. Tang, T. H. Cheng, H. Y. Tam, and C. Lu, Opt. Lett. **32**, 1581 (2007).

## Figure captions:

Fig.1: Schematic of a gain-guided dispersion-managed fiber laser. λ/4: quarter-wave plate; PI: polarization dependent isolator; λ/2: half-wave plate; DCF: dispersion compensated fiber; WDM: wavelength division multiplexing; EDF: erbium-doped fiber.

Fig. 2: (a)/(d) Optical spectrum; (b)/(e) Autocorrelation trace; (c)/(f) Oscilloscope trace of a typical multiple-pulsing state of two GGSs with large spacing / a bound state of two GGSs.

Fig. 3: Numerically simulated bound state of two GGSs (solid: at the position with minimum pulse width; dash: at the position with maximum pulse with; linear cavity phase delay bias is 1.85π, pump strength is 2800).

Fig. 4: Pulse width evolution of the numerically obtained bound state of GGSs in the fiber laser.

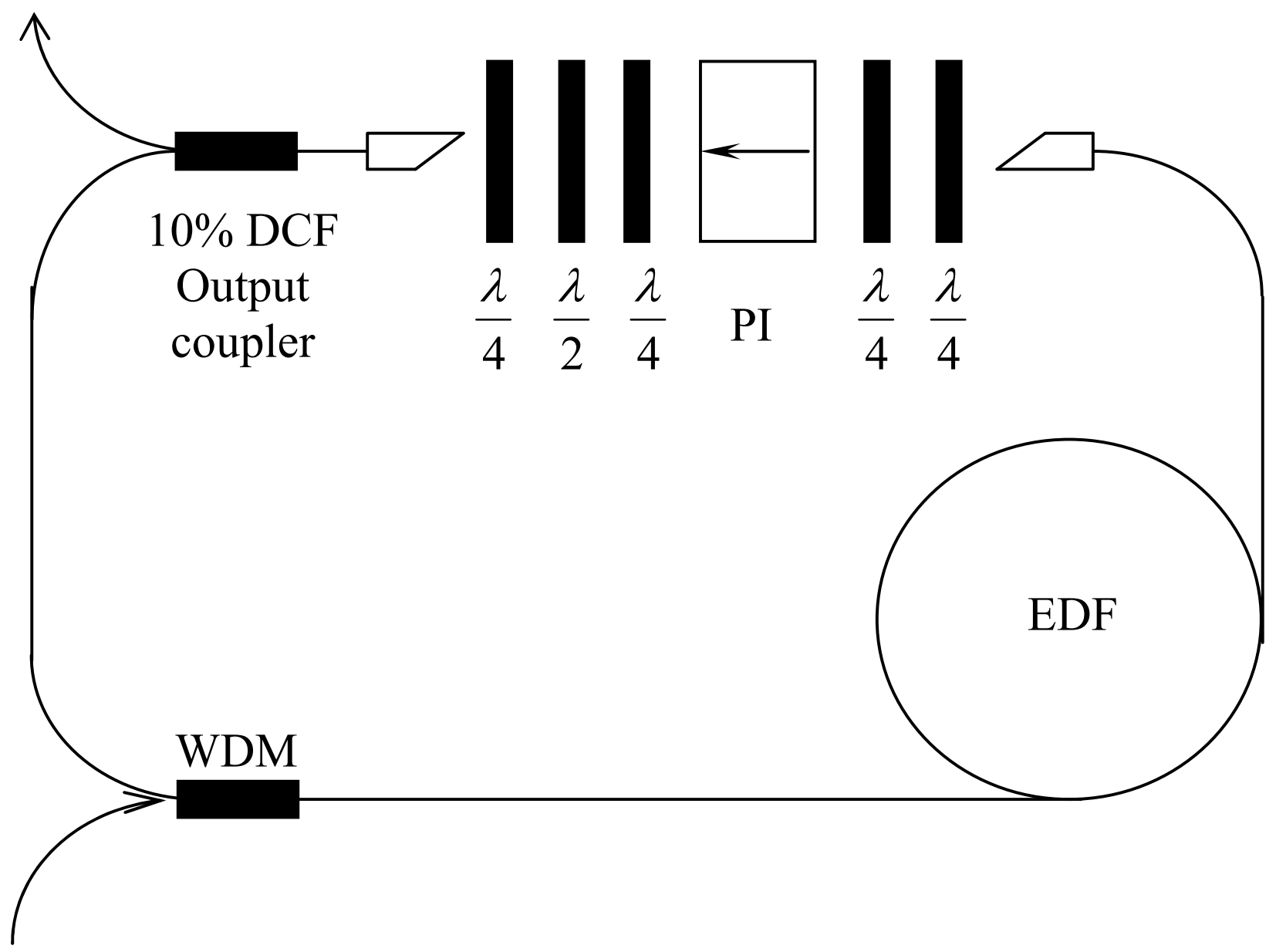

Fig. 1 L. M. Zhao et al.

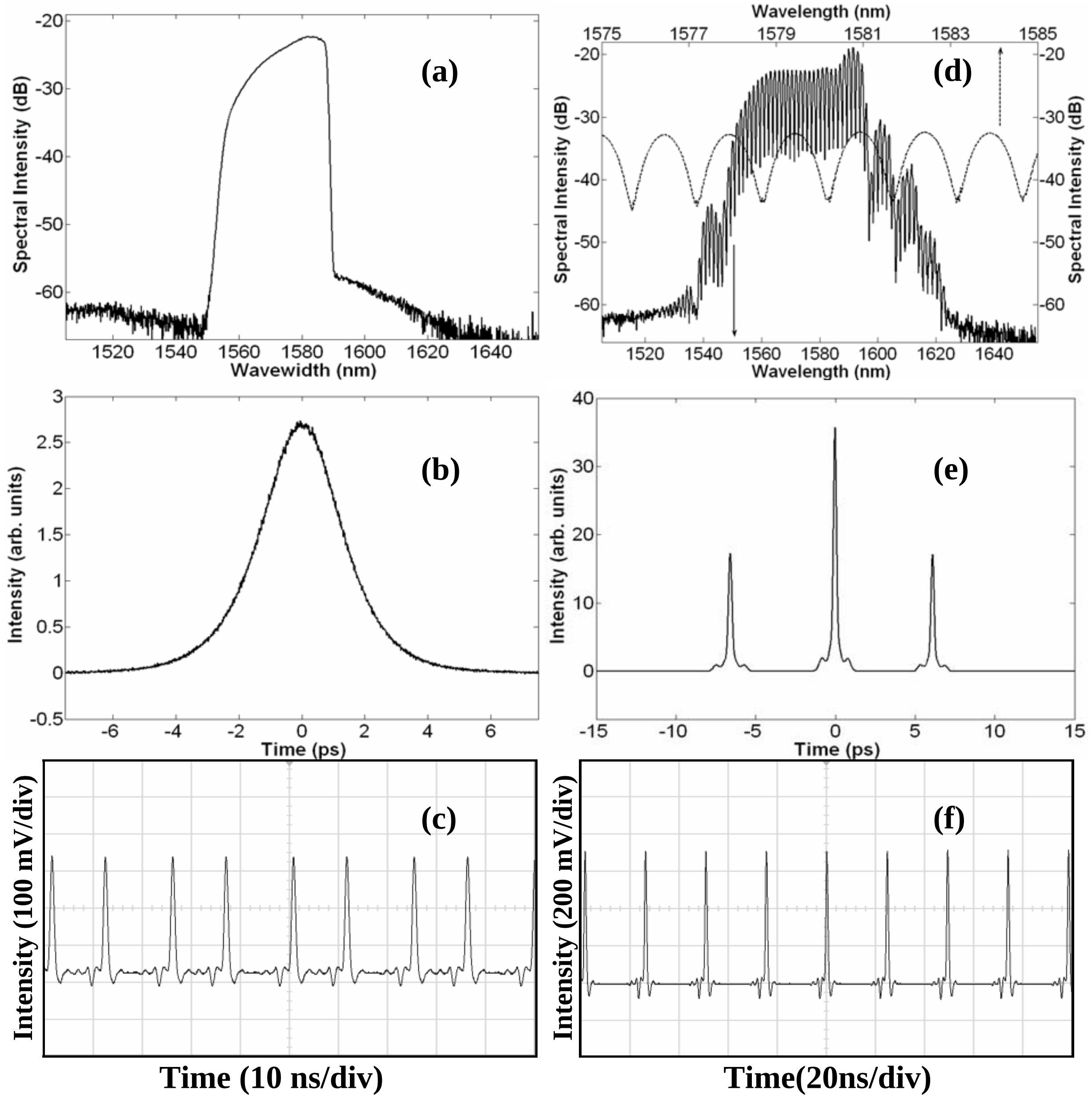

Fig. 2 L. M. Zhao et al.

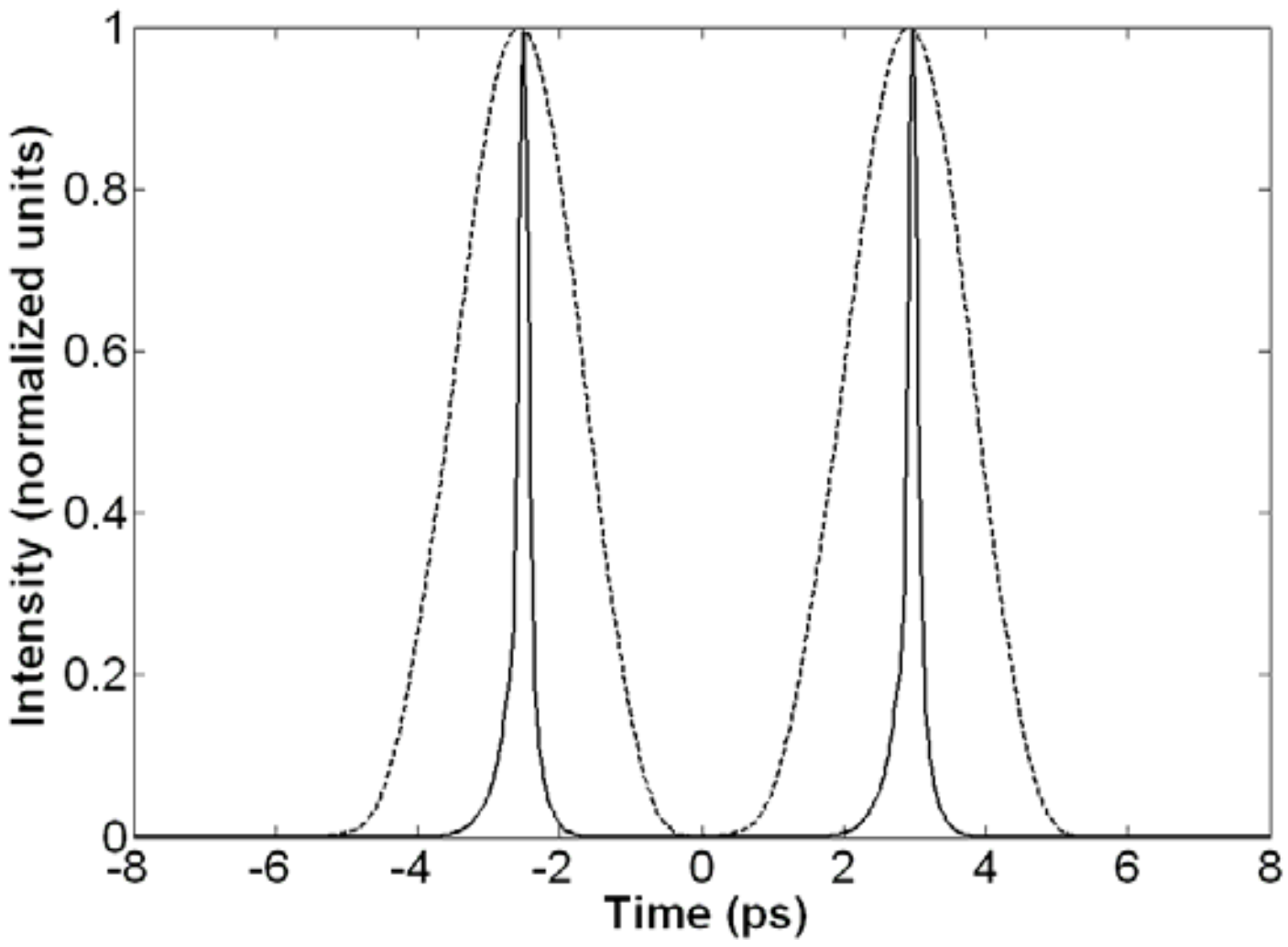

Fig. 3   L. M. Zhao et al.

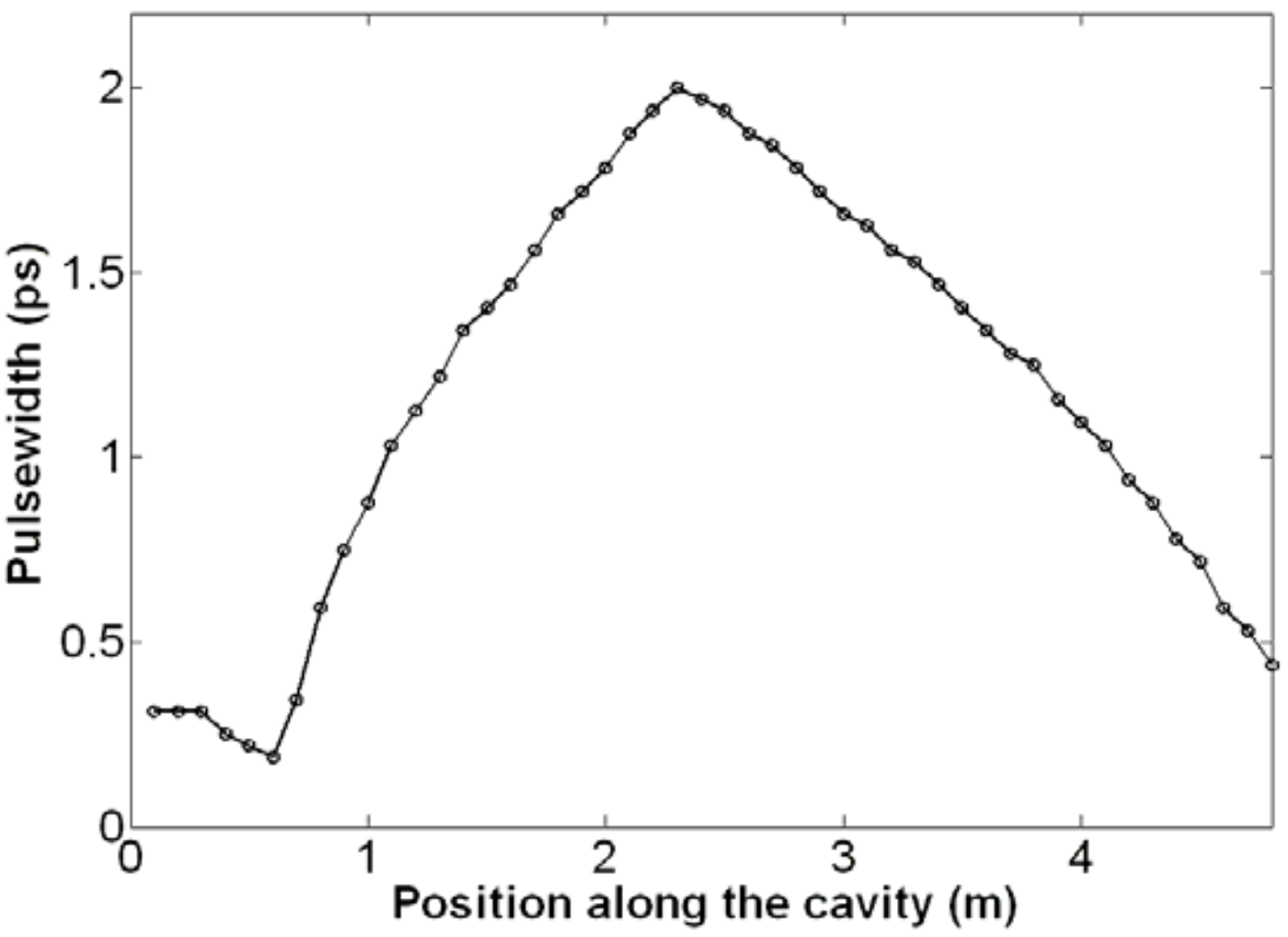

Fig. 4 L. M. Zhao et al.